\documentclass[journal,onecolumn]{IEEEtran}

\usepackage{amsmath,amsxtra,amssymb,latexsym, amscd,amsthm}
\usepackage[mathscr]{eucal}
\usepackage{url}
\usepackage{algorithm, algorithmic}

\usepackage{color} 
\newtheorem{theorem}{Theorem}
\newtheorem{definition}{Definition}
\newtheorem{lemma}{Lemma}
\newtheorem{proposition}{Proposition}
\newtheorem{remark}{Remark}
\newtheorem{example}{Example}
\newtheorem{corollary}{Corollary}

\newcommand{\ie}{{\em i.e. }}

\newcommand{\Bs}{\mathscr{B}}
\newcommand{\Bc}{\mathcal{B}}
\newcommand{\B}{\Bc}

\newcommand{\Cc}{\mathcal{C}}
\newcommand{\Cf}{\mathfrak{C}}

\newcommand{\E}{{\mathbb{E}}}
\newcommand{\Ec}{{\mathcal{E}}}

\newcommand{\F}{{\mathbb{F}}}

\newcommand{\Gc}{\mathcal{G}}

\newcommand{\Lc}{\mathcal{L}}

\newcommand{\Mc}{\mathcal{M}}

\newcommand{\Hc}{\mathcal{H}}

\everymath{\displaystyle}

\title{Generalized subspace subcodes with application in cryptology}

\author{Thierry~P.~BERGER, Cheikh~Thi\'ecoumba~GUEYE and Jean~Belo~KLAMTI}

\date{\today}

\begin{document}
\maketitle

\thanks{Cheikh~Thi\'ecoumba~GUEYE and Jean~Belo~KLAMTI are with Universit\'e Cheikh Anta Diop, 
Facult\'e des Sciences et Techniques, DMI, LACGAA, Dakar, S\'en\'egal, 
\url{cheikht.gueye@ucad.edu.sn} and \url{jeanbelo.klamti@ucad.edu.sn}
}

\thanks{Thierry~P.~Berger is with XLIM
(UMR~CNRS~6172), Universit\'e de Limoges, 123 avenue A. Thomas, 87060 Limoges
Cedex, France,  \url{thierry.berger@unilim.fr}
}

\begin{abstract}
   
 Most of the codes that have an algebraic decoding algorithm are derived from the Reed Solomon codes.
 They are obtained by taking equivalent codes, for example the generalized Reed Solomon codes, or by 
 using the so-called subfield subcode method, which leads to Alternant codes and Goppa codes over the 
 underlying prime field, or over some intermediate subfield. The main advantages of these constructions is 
 to preserve both the minimum distance and the decoding algorithm of the underlying Reed Solomon code.
 In this paper, we propose a generalization of the subfield subcode construction by introducing the notion of
 subspace subcodes and a generalization of the equivalence of codes which leads to the notion of generalized 
 subspace subcodes. When the dimension of the selected subspaces is equal to one, we  show that our approach 
 gives exactly the family of the codes obtained by equivalence and subfield subcode technique. However, our 
 approach highlights the links between the subfield subcode of a code defined over an extension field and the 
 operation of puncturing the $q$-ary image of this code.
 When the dimension of the subspaces is greater than one, we obtain codes whose alphabet is no longer a 
 finite field, but a set of $r$-uples. We explain why these codes are practically as efficient for applications 
 as the codes defined on an extension of degree $r$. In addition, they make it possible to obtain decodable 
 codes over a large alphabet having parameters previously inaccessible. 
 As an application, we give some examples that can be used in public key cryptosystems such as McEliece.

\end{abstract}
	
\begin{IEEEkeywords}
Linear code, Shortened code, Punctured code, Subfield subcodes, Reed Solomon codes, Alternant codes, $q$-ary image.
\end{IEEEkeywords}

% \tableofcontents
 \section{Introduction}

 %%%%%%%%%%%%%%%%%%%

The McEliece cryptosystem is the most known and oldest code-based cryptographic protocol.
An important part of its security is based on the use of codes that seem random and possess an 
effective error correction algorithm. In its original paper, R. McEliece proposed the use of  binary Goppa 
codes. This class is a subclass of Alternant codes, which are themselves subcodes on the binary field of 
Generalized Reed-Solomon codes. This construction makes it possible to easily decode errors, provides a 
good minimum distance and effectively mask the underlying algebraic structure.

The main problem with this protocol is the size of the secret key. There are several ways to reduce the 
size of keys. One of these is the use of codes with a large automorphism group, typically quasi-cyclic (QC),
quasi-dyadic (QD), or quasi-monoidic (QM) matrices \cite{BCG,BL,Gab,MB1,MBTS,MBL,Per}.

Another approach is to use subfield subcodes over a subfield of great size.
The variant based on the subfield subcodes introduced by \textit{Berger et al.} \cite{BL} was attacked by
\textit{Wieschebrink} \cite{Wies}.
Recently \textit{Faug\`{e}re et al.} proposed two attacks respectively a structural attack and a algebraic
attack against the McEliece schemes with compact keys \cite{FOPPT,FOPT}.

In this paper, we introduce a new construction of subfield subcodes called \textit{Generalized Subfield
Subcodes} and we prove that the \textit{Generalized Subfield Subcodes} of Reed-Solomon are exactly alternant
codes.
The approach developed for the \textit{Generalized Subfield Subcodes} leads to a second construction called
\textit{Generalized Subspace Subcodes} which is a promising research direction for both coding theory and
hiding the structure of a code.

This paper is organized as following: in Section \ref{sec:2} we give some definitions in coding theory.
In Section \ref{sec 2} we introduce the shortened $q$-ary images of a code and give the link between subfield
subcodes and shortened codes.
In Section \ref{sec: 3} we present the first attempt at generalization of subfield subcodes namely the
\textit{Generalized Subfield Subcodes} and we show that the codes introduced in Section
\ref{sec:first-construction} can also be constructed using a known method to construct alternant codes.
In Section \ref{sec:subspacesubcode} we introduce \textit{Subspace Subcodes}, which is a new class of additive
block codes.
We generalize this first class to obtain another class of block additive codes named \textit{Generalized
Subspace Subcodes}.
For this second class we proposed an algorithm which allows us to compute its generator matrix.
In addition we give some examples and directions for their application in transmission and cryptology.

\section{Preliminaries}\label{sec:2}
\subsection{Linear code}
Let $\F_{q^m}$ be an arbitrary finite field.  A linear code $\Cc$ of length $n$ and dimension $k$ is a vector
subspace of $\F_{q^m}^n$ of dimension $k$.  A vector $x\in \F_{q^m}^n$ is called word and a vector $x\in
\Cc$ is called codeword.

The Hamming distance between two words $x$ and $y$ denoted by $d(x, y)$ is the number of positions on which
they differ.
The Hamming distance of a code $\Cc$ denoted by $d$ is the minimal Hamming distance between any two different
codewords.

The Hamming weight of a word $x\in \F_{q^m}^n$ denoted by $wt(x)$ is the number of its nonzero coordinates.
In the case of a linear code the minimal Hamming distance of a code is equal to the minimal Hamming weight of
its
nonzero codewords.

A linear code $\mathcal{C}$ over an arbitrary finite field $\F_{q^m}$ is called $\mathbb{F}_{q^m}$-linear
code.  If its length is $n$, its dimension is $k$ and its minimal Hamming distance is $d$ we call this code a 
$[n, k, d]$ $\F_{q^m}$-linear
code.    A linear code
$\mathcal{C}$ over an arbitrary finite field $\F_{q^m}$ is usually specified by a full-rank matrix
$\mathcal{G}\in \F_{q^m}^{k\times n}$ called \textit{generator matrix} of $\mathcal{C}$, whose rows span the
code.  Thus, $\mathcal{C}=\left\lbrace x\mathcal{G} \\: x\in \F_{q^m}^{k} \right\rbrace $.  A linear code can
be also defined by the right kernel of a matrix $\mathbf{H}$ called \textit{parity-check matrix} of
$\mathcal{C}$ as follows:
\[ 
\mathcal{C}=\left\lbrace x\in \F_{q^m}^{k}  \ \ s.t. \ \  \mathbf{H}x^{T}=0 \right\rbrace 
\]

The matrix $H$ is a generator matrix of the dual code $\Cc^{\bot}$ of $\Cc$ for the usual scalar product.

\subsection{Shortened codes and punctured codes}

\begin{definition}(Shortened code)\cite{CG}

Let $\Cc$ be an $[n,k,d]_{q^m}$-linear code, choose a subset $I\subset \left\lbrace 1, 2, ...,n \right\rbrace
$
of coordinates such that $|I|=i$ with $1\leq i\leq n$ and take the subcode of $\Cc$ consisting of the
codewords having $0$ on those positions.  Deleting the chosen coordinates in every codeword of this subcode
yields a $\mathbb{F}_{q^m}$-linear code denoted $Short_I(\mathcal{C})$.  $Short_I(\mathcal{C})$ is
called a shortened code of $\mathcal{C}$ on $I$.
\end{definition}

\begin{definition}(Punctured code)\cite{CG}\label{Pun}

Let $\Cc$ be an $[n,k,d]_{q^m}$-linear code,choose a subset $I\subset \left\lbrace 1, 2, ...,n \right\rbrace $
of coordinates such that $|I|=i$ with $1\leq i<d$.  Deleting the chosen coordinates in every codeword yields
a 
$\mathbb{F}_{q^m}$-linear code denoted $Punct_I(\mathcal{C})$.  $Punct_I(\mathcal{C})$ is called a 
punctured code of $\mathcal{C}$ on $I$.
\end{definition}

If $I=\left\lbrace j\right\rbrace $, we denote $Punct_I(\Cc)$ by $Punct_j(\Cc)$ and $Short_I(\Cc)$ by
$Short_j(\Cc)$ where $j$, $1\leq j\leq n$, is the deleted position.

For all vector $x\in \F_{q^m}^n$, we denoted by $Punct_I(x)=(x_i)_{i\notin I}$, if $x$ is such that
$x_I=(x_i)_{i\in I}=0$, we denoted by $Short_I(x)=(x_i)_{i\notin I}$.

\begin{remark}\label{rem:Racc}
There are some links between shortening and  puncturing operations.  Indeed, let $\Cc$ be an
$[n,k,d]_{q^m}$-linear code.  
Let $I$ be a subset of $N=\left\lbrace 1,2,..., n \right\rbrace $.  The shortened code of $\Cc$ on $I$ is the
punctured code 
on $I$ of the subcode 
\\
$\Cc_I=\left\lbrace \textbf{c}=(c_1, c_2,..., c_n)\in \Cc \: | \: c_i=0\ \forall i\in I\right\rbrace $.  We
remark also that a 
shortened code of a linear code $\Cc$ can be considered like a subcode of $\Cc$ if we replace
the deleted coordinates by $0$.  Therefore, all the best decoding algorithms of $\Cc$, can be used to decode a
shortened code of $\Cc$.
\end{remark}

\begin{theorem}\label{thm:puncrac}
	Let $\Cc$ be an $[n,k,d]$ $\F_{q^m}$-linear code.  Let $I$ be a subset of $N=\left\lbrace
	1,2,...,n\right\rbrace $.  Then the following identity is verified
$$
Punct_I \left( \Cc\right)^{\bot} =Short_I\left( \Cc^{\bot} \right)
$$
\end{theorem}

\begin{IEEEproof}

Let $x\in \Cc$, $y\in \Cc^{\bot}$.  Then if $(y_i)_{i\in I}=0$, we have 
$Short_I(y)=(y_k)_{k\not\in I}\in Short_{I}(\Cc^{\bot})$ and $Punct_I(x)=(x_k)_{k\not\in I}\in
Punct_{I}(\Cc)$.  
According to the definition of a code and its dual we have
$$
x.y^T=0\Longleftrightarrow \sum\limits_{k\in N}x_ky_k =\sum\limits_{k\in N\backslash I}x_ky_k
+\sum\limits_{k\in I}x_ky_k =0
$$
$$
\hspace{1.8cm}\Longleftrightarrow \sum\limits_{k\in N\backslash I}x_ky_k=Punct_I(x).Short_I(y)^{T}=0.
$$

\end{IEEEproof}

\begin{lemma}\label{lem: lem1}
	Let $\Cc$ be an $[n,k,d]$ $\F_{q^m}$-linear code.  Let $i \in\{1, ..., n\}$.  The equality $Short_i( \Cc)
	=Punct_i( \Cc)$ is verified if and only if one of the following conditions is satisfied:
	\begin{enumerate}
		\item For all codewords $c=(c_1, c_2,...,c_n)\in\Cc$, $c_i=0$, 
		\item The word
		$e_i=(0,...,0,1,0,...,0)$ having only one non-zero coefficient which is equal to 1 in position $i$ is in
$\Cc$.
	\end{enumerate}
\end{lemma}

\begin{IEEEproof}
\begin{enumerate}
	\item Suppose first that the identity $Short_i( \Cc) =Punct_i( \Cc)$ is verified.

	 Suppose that Conditions 1) is not satisfied, then there exists a codeword $c\in \Cc$ such that $c_i=1$.  
                      Under our hypothesis, $Punct_i(c)$ is an element of $Short_i( \Cc)$, \ie there exists
$c'\in \Cc$ such that $P00unct_i(c)=Short_i(c')$.
	Clearly, $e_i=c'-c$ is an element of $\Cc$ and Condition 2) is satisfied.
	 
	\item Reciprocally
	\begin{enumerate}
		\item Suppose that Condition 1) is satisfied.  Since all the codewords $c \in \Cc$ verify
		$c_i=0$, then $Punct_i(\Cc)=Short_i(\Cc)$.  \item Suppose that Condition 2) is satisfied.  Let
		$\Cc_i=\left\lbrace c\in \Cc\ \ s.t\ \ c_i=0 \right\rbrace$ be the subcode of $\Cc$
		constituted of codewords $c$ such that $c_i=0$.  Clearly, $\Cc$ is generated by $\left\lbrace
		e_i\right\rbrace \cup \Cc_i$.  Let $c$ be a codeword of $\Cc$.  If $c\in \Cc_i$ then
		$Punct_i(c)=Short_i(c)\in Short_i(\Cc)$.  If $c=e_i+c'$, $c'\in \Cc_i$, then
		$Punct_i(c)=Short_i(c')\in Short_i(\Cc)$.
	\end{enumerate}
	
	In both cases, $Short_i( \Cc) =Punct_i( \Cc)$.
\end{enumerate}
\end{IEEEproof}

\begin{remark}
   If a code $\Cc$ satisfies the first condition of Lemma \ref{lem: lem1}, then its dual $\Cc^{\bot}$ will
   satisfy the second one.
\end{remark}

We deduce the following proposition.

\begin{proposition}\label{pro: p1}
Let $\Cc$ be an $[n,k,d]$ $\F_{q^m}$-linear code and $i$, $1\leq i\leq n$, be an integer.  If the parameters
of
$Punct_i(\Cc)$ and $Short_i(\Cc)$ are respectively $[n-1,k_p,d_p]$ and $[n-1,k_s,d_s]$, then:
 \begin{enumerate}
    \item $d_s \geq d$, $ d_s \geq d_p\geq d-1 $.
    
    \item If $Punct_i(\Cc) \neq Short_i(\Cc)$ then $k_p=k$ and $k_s=k-1$.
    
    \item If $Punct_i(\Cc) = Short_i(\Cc)$ then
    \begin{itemize}
        \item If Condition $1$ of \textit{Lemma} \ref{lem: lem1} is verified, then the parameters of
        $Punct_i(\Cc)$ and $Short_i(\Cc)$ are $[n-1,k,d]$.  \item If Condition $2$ of \textit{Lemma} \ref{lem:
        lem1} is verified (\ie $e_i \in \Cc$), then $k_p=k_s=k-1$.
  	\end{itemize}
  \end{enumerate}
\end{proposition}

\begin{IEEEproof}
   
   Note that, since $Short_i(\Cc)\subset Punct_i(\Cc)$, the following relations hold: $k_s \leq k_p \leq k$
   and $d_s\geq d_p$.  Moreover, using the notations of the proof of Lemma \ref{lem: lem1}, $Short_i(\Cc)$ is
   isomorphic to $\Cc_i \subset \Cc$, and then $d_s\geq d$.
   
   One can easy check that $d_p\geq d-1$.
   
   Suppose firstly that for all codewords $c=(c_1, c_2,...,c_n)\in\Cc$, $c_i=0$ (Condition $1$ of
\textit{Lemma} \ref{lem: lem1}). Then we have, $Punct_i(\Cc) = Short_i(\Cc)$ is an $[n-1,k,d]$ code.
   
   Suppose now that there exists a codeword $c \in \Cc$ such that $c_i=1$.  One can check that the code $\Cc$
   is equal to $\left\langle \left\lbrace c\right\rbrace \right\rangle \oplus \Cc_i$.  We deduce that
   $k_s=k-1$.
   
   If $e_i \not\in \Cc$, then $Short_i(\Cc)\subsetneq Punct_i(\Cc)$, and then $k_s=k-1 < k_p \leq k$.
   
   If $e_i \in \Cc$, then $k_p=k_s=k-1$.

\end{IEEEproof}

One can notice that if $e_i \in \Cc$, then $d=1$, and we have no information about values of $d_s$ and $d_p$
(but we have in this case $d_s=d_p$).
\\
From Proposition \ref{pro: p1}, we deduce the following corollary:

\begin{corollary}\label{cor:1}
Let $\mathcal{C}$ be an $[n, k, d]$ $\F_{q^m}$-linear code and $I$ be a set of $r$ distinct positions.  Let
$[n-1,k_s,d_s]$ and $[n-1,k_p, d_p]$ be respectively the parameters of $Short_I(C)$ and $Punct_{I}(C)$.  Then
we have $ d_s\geq d$, $ k_s\geq k-r$, $ d_p \geq d-r$ and $ k_p\geq k-r$.
\end{corollary}

\subsection{Subfield Subcodes and Trace code}\label{sec:sfsc} 

For more details and proofs, the reader can refer to \cite{McW}, Ch.7 \S 7.

\begin{definition}
   The subfield subcode $\mathfrak{C}$ over $\F_q$ of a $\F_{q^m}$-linear code $\Cc$ is the set of codewords
of $\Cc$ 
   which have components in $\F_q$:
   $\mathfrak{C}=\Cc\cap \F_q^{n}$.
\end{definition}

A first property of $\mathfrak{C}$ is the fact that it is a $\F_q$-linear code. The simplest way to construct 
such a subfield subcode is to construct a parity check matrix as follows.

Let $\Cc$ be an $[n,k,d]$ $\F_{q^m}$-linear code defined by the parity check matrix 
$$\mathbf{H}=\begin{pmatrix} h_{1,1}& ...  & h_{1,n}
\\
\colon& & \colon \\
h_{r, 1}& ...  & h_{r, n} \end{pmatrix} \in \F_{q^m}^{r\times n}.$$

Let $ \Bc=\left\lbrace b_1, b_2, ..., b_m\right\rbrace$
 be a basis of $\F_{q^m}$ as a $\F_{q}$-vector space. We can construct the map 
 $\phi_{\Bc}: \F_{q^m}\longrightarrow \F_{q}^{m}$ defined by, if $x=\sum_{i=1}^{m}x_i b_i$, 
 $x_i \in \F_q$, 
 then $\phi(x)=(x_1,x_2, ...,x_m)$.
 
 \begin{proposition}\label{prop2}
  The matrix  $
\tilde{\mathbf{H}}=\begin{pmatrix} \phi(h_{1,1})^T& ...  & \phi(h_{1,n})^T \\
\colon& & \colon \\
\phi(h_{r,1})^T& ...  & \phi(h_{r,n})^T
\end{pmatrix}
$ is a parity check matrix of the subfield subcode $\Cf$ of $\Cc$.
 \end{proposition}
 
 Note that $\tilde{\mathbf{H}}$ is not necessary of full rank. Then a subfield subcode $\Cf$ of an 
 $[n,k,d]$ $\F_{q^m}$-linear code $\Cc$ is an $[n, k^*\geq n-rm, d^*\geq d]$ $\F_q$-linear
code.

One can notice that $\tilde{\mathbf{H}}$ is independent of the choice of the basis $\Bc$, which allows to omit
the index $\B$ in the definition of $\phi_{\Bc}$.

\begin{example}
The subfield subcode of the Reed-Solomon code $RS_d$ of minimal distance $d$ 
is the BCH code $BCH_d$ of constructed minimal distance $\delta=d$ over the
prime subfield $\F_p$. Note that the true minimum distance of $BCH_d$ could be greater than $d$.
\end{example}

Another construction of a $\F_q$-linear code from a $\F_{q^m}$-linear code is the trace construction.
\\
If $x$ is an element of $\F_{q^m}$, the trace of $x$ over $\F_q$ is defined by $T_m(x)=x+x^{q}+x^{q^2}+\cdots 
+x^{q^{m-1}}$.  The trace function is a $\F_q$-linear map.
This mapping is naturally extended to $\F_{q^m}^n$: if $c=(c_1,...,c_n)\in \F_{q^m}^n$, then 
$T_m(c)=(T_m(c_1),...,T_m(c_n))\in \F_q^n$.

\begin{definition}\cite{McW}
The trace code of an $\F_{q^m}$-linear code $\Cc$ is the $\F_q$-linear code $T_m(\Cc)$.
\end{definition}

The link between trace code and subfield subcode is described in the following theorem:
\begin{theorem}\cite{McW}[Th.11, ch.7 \S7, Delsarte]
The dual of the subfield code
$\Cf$ of a code $\Cc$ is the trace code of its dual: 
\\ \phantom{a}\hspace{16mm}
$\Cf^{\bot}=T_m(\Cc^{\bot})$.
\end{theorem}

This fact is a direct consequence of Proposition \ref{prop2} and a classical result of algebra: all
$\F_q$-linear mapping of $\F_{q^m}$ into $\F_q$ can be expressed as $T_{m}(\alpha x)$ for some $\alpha \in
\F_{q^m}$.

\subsection{$\F_{q^m}$-linear isometries and Alternant codes} \label{automorphism}

It is well-known \cite{Huf} that the linear  isometries for the Hamming distance on $\F_{q^m}^n$ form a group

generated by the permutations of the support and the scalar multiplications by invertible elements of
$\F_{q^m}$ 
on each coordinate. From a matrix point of view, it is the monomial group $\Mc_n$ of $n\times n$ 
matrices over $\F_{q^m}$ with one and only one non-zero element on each row and each column. 

In order to obtain a code equivalent to $\Cc$, such a monomial matrix acts by right multiplication on any
generator 
matrix of a code $\Cc$. The new code is $\F_{q^m}$-linear and has the same parameters  of the original one.

Moreover, if $\Cc$ has a decoding algorithm, then this algorithm can be used to decode the new code.

The most famous example is that of Generalized Reed-Solomon (GRS) codes that are obtained by applying a
monomial matrix to a Reed-Solomon code.

An Alternant code is simply a subfield subcode of a GRS code. 
It naturally inherits the decoding algorithm of the underlying Reed-Solomon code.

\subsection{$q$-ary images of a code of length $n$ over $\F_{q^m}$}\label{section:qary image}

As we did in Section \ref{sec:sfsc}, we fix a basis $\Bc=(b_1,...,b_m)$ of $\F_{q^m}$ over $\F_q$ and denote 
by $\phi_{\Bc}$ the corresponding $\F_q$-linear isomorphism $\F_{q^m}\mapsto \F_q^m$. 

The mapping $\phi_{\Bc}$ can be extended to the whole space $\F_{q^m}^n$: if $c=(c_1,...c_n) \in \F_{q^m}^n$, 
then $\Phi_{\Bc}(c)=(\phi_{\Bc}(c_1),...,\phi_{\Bc}(c_n))$.

\begin{definition}
The $q$-ary image of a code  $\Cc$ relative to the base $\Bc$ is the
image $Im_q(\Cc)=\Phi_{\Bc}(\Cc)$ of $\Cc$ by $\Phi_{\Bc}$.
\end{definition}

The code $Im_q(\Cc)$ is clearly a $\F_q$-linear code of length $nm$.  Note that, contrary to Section
\ref{sec:sfsc}, this code is dependent on the choice of the basis $\Bc$.

In order to build a generator matrix $G$ of $Im_q(\Cc)$ over $\mathbb{F}_q$ from those $\Gc$ of $\Cc$ over
$\mathbb{F}_{q^m}$, since $Im_q(\Cc)$ is not $\F_{q^m}$-linear, it is necessary to take all the multiples of
the rows of $\Gc$.  In fact, it is sufficient to take $m$ multiples $\F_q$-linearly independent.

A simple way is the following: for any element $\beta \in \F_{q^m}$, the map $\psi_{\beta}$: $x \mapsto \beta
x$ is a $\F_q$-linear endomorphism of $F_{q^m}$.  Its image by $\phi_{\Bc}$ is an endomorphism of $\F_q^m$.
We denote by $M_{\beta}$ the matrix of the corresponding endomorphism: with obvious notations, if
$\phi_{\Bc}(x)=(x_1,...,x_m)$ then $\phi_{\Bc}(\beta x)=(x_1,...,x_m)M_{\beta}$.

\begin{proposition}

If $\Gc=(\beta_{i,j})$ is a $k\times n$ generator matrix of $\Cc$, then the $mk \times nm$ matrix $G$ obtained
by replacing each entry $\beta_{i,j}$ by the corresponding $m \times m$ matrix $M_{\beta_{i,j}}$.  Moreover,
the matrix $G$ is of full rank $km$.
\end{proposition}

\begin{IEEEproof}
   The fact that $G$ generates the full code $Im_{\Bc}(\Cc)$ comes directly from the fact that $\Phi_{\Bc}$ 
   is an isomorphism. In addition the two codes have the same number of elements, which implies that $G$ is
of 
   rank $km$.
\end{IEEEproof}

As a direct consequence, we obtain the following corollary:

\begin{corollary}\label{cor:2}
If $\Cc$ is an $[n,k,d]$ $\F_{q^m}$-linear code, then $Im_q(\Cc)$ is an $[nm, km, d_q\geq d]$ $\F_q$-linear
code.
\end{corollary}

\section{Link between Subfield Subcodes and Shortened codes}\label{sec 2}

\subsection{Shortening the $q$-ary image of a code}\label{sec:first-construction}

Let $u=(i_1, i_2,...,i_n )\in \left\lbrace 1, 2, ..., m \right\rbrace^n$ be a $n$-tuple of positions $i_j$,
$1\leq i_j \leq m$.  We define two sets of indexes \\
$I_{u}=\left\lbrace i_1, i_2+m, i_3+2m; ..., i_n+(n-1)m \right\rbrace$ and 
$J_u=\overline{I_u}= \{1,...,n\}\setminus I_{u}$.

Let $\Cc$ be a linear code of length $n$ over $\F_{q^m}$. We fix a basis $\Bc$ of  $\F_{q^m}$ over $\F_q$ and 
we look at the $q$-ary image of $\Cc$ relatively to $\Bc$.

We denote by $S_{u}$ (respectively $P_{u}$) the operation of shortening (respectively puncturing) the  
$q$-ary image of $\Cc$ on positions $J_u$:
$S_u(\Cc)=Short_{J_u}(Im_q(\Cc))$  and $P_u(\Cc)=Punct_{J_u}(Im_q(\Cc))$.

\begin{proposition}
If $\Cc$ is an $[n,k,d]$ $\F_{q^m}$-linear code, then $S_u(\Cc)$ is an $[n,k',d']$ $\F_{q}$-linear code with
$k' \geq
n-m(n-k)$ and $d' \geq d$.
Moreover, if the code $\Cc$ has a decoding algorithm of error correction capability $t$, then this algorithm
can be 
applied to $S_u(\Cc)$ with the same error correction capability.
\end{proposition}

\begin{IEEEproof}
The inequalities $k' \geq n-m(n-k)$ and $d' \geq d$ are direct consequences of Corollary \ref{cor:1} and  
Corollary \ref{cor:2}. In order to decode a noisy codeword $y$ of $S_u(\Cc)$, we extend $y$ to a word of 
length $nm$ by adding the value 0 on the shortened position, then we use the inverse of the map $\Phi_{\Bc}$ 
in order to obtain a noisy codeword $\mathbf{y}$ of $\Cc$. By construction, the weight of the errors on $y$ 
and $\mathbf{y}$ are the same. So, if the error is less than or equal to $t$, it is possible to correct
$\mathbf{y}$ 
and to recover the correct codeword $c\in S_u(\Cc)$.
\end{IEEEproof}

\begin{example} \label{example:1}
Set $n=7$, $m=3$ and let $\alpha$ be a root of the polynomial $x^3+x+1$. The following matrix is a generator 
matrix of the Reed Solomon code $RS_2$ of parameters $[7,6,3]_8$ associated to the support 
$a=(1,\alpha, \alpha^2, \alpha^3, \alpha^4, \alpha^5, \alpha^6)$:
$$
\mathcal{G}=\begin{pmatrix}
1 & 1       & 1        & 1       & 1      & 1      & 1      \\ 
1 &  \alpha & \alpha^2 & \alpha^3 & \alpha^4 & \alpha^5 & \alpha^6 \\ 
1 & \alpha^2 & \alpha^4 & \alpha^6 & \alpha & \alpha^3 & \alpha^5 \\
1 & \alpha^3 & \alpha^6 & \alpha^2 & \alpha^5 & \alpha & \alpha^4 \\ 
1 &  \alpha^4 & \alpha & \alpha^5 & \alpha^2 & \alpha^6 & \alpha^3 \\ 
1 &  \alpha^5 & \alpha^3 & \alpha & \alpha^6  & \alpha^4 & \alpha^2 \\ 
\end{pmatrix}
$$

Its generator matrix in form systematic is given by:
$$
\mathcal{G}_{sys}=\begin{pmatrix}
1 & 0         & 0        & 0        & 0        & 0        & \alpha   \\ 
0 & 1         & 0        & 0        & 0        & 0        & \alpha^2 \\ 
0 & 0         & 1        & 0        & 0        & 0        & \alpha^3 \\
0 & 0         & 0        & 1        & 0        & 0        & \alpha^4 \\ 
0 & 0         & 0        & 0        & 1        & 0        & \alpha^5 \\ 
0 & 0         & 0        & 0        & 0        & 1        & \alpha^6 \\ 
\end{pmatrix}
$$
The $q$-ary image (binary image) of the generator matrix $\mathcal{G}$ in the base $\left\lbrace 1=(100), \
	\alpha=(010), \ \alpha^2=(001)\right\rbrace $
	is given by
	$$
	Im_2\left( \mathcal{G}_{syst}\right) =\begin{pmatrix}
	M_{1} & M_{0} & M_{0} & M_{0} & M_{0} & M_{0} & M_{\alpha} \\ 
	M_{0} &  M_{1} & M_{0} & M_{0} & M_{0} & M_{0} & M_{\alpha^2} \\ 
	M_{0} & M_{0} & M_{1} & M_{0} & M_{0} & M_{0} & M_{\alpha^3} \\
	M_{0} &  M_{0} & M_{0} & M_{1} & M_{0} &M_{0} & M_{\alpha^4} \\ 
	M_{0} &  M_{0} & M_{0} & M_{0} & M_{1} & M_{0} & M_{\alpha^5} \\
	M_{0} &  M_{0} & M_{0} & M_{0} & M_{0} & M_{1} & M_{\alpha^6} \\ 
	\end{pmatrix}
	$$
	then
	$$
	Im_2(\mathcal{G})=\left( \begin{array}{ccc|ccc|ccc|ccc|ccc|ccc|ccc}
	1 & 0 & 0 & 0 & 0 & 0 & 0 & 0 & 0 & 0 & 0 & 0 & 0 & 0 & 0 & 0 & 0 & 0 & 0 & 1 & 0 \\ 
	0 & 1 & 0 & 0 & 0 & 0 & 0 & 0 & 0 & 0 & 0 & 0 & 0 & 0 & 0 & 0 & 0 & 0 & 0 & 0 & 1 \\ 
	0 & 0 & 1 & 0 & 0 & 0 & 0 & 0 & 0 & 0 & 0 & 0 & 0 & 0 & 0 & 0 & 0 & 0 & 1 & 1 & 0
	\\
	\hline
	0 & 0 & 0 & 1 & 0 & 0 & 0 & 0 & 0 & 0 & 0 & 0 & 0 & 0 & 0 & 0 & 0 & 0 & 0 & 0 & 1 \\ 
	0 & 0 & 0 & 0 & 1 & 0 & 0 & 0 & 0 & 0 & 0 & 0 & 0 & 0 & 0 & 0 & 0 & 0 & 1 & 1 & 0 \\ 
	0 & 0 & 0 & 0 & 0 & 1 & 0 & 0 & 0 & 0 & 0 & 0 & 0 & 0 & 0 & 0 & 0 & 0 & 0 & 1 & 1
	\\
	\hline
	0 & 0 & 0 & 0 & 0 & 0 & 1 & 0 & 0 & 0 & 0 & 0 & 0 & 0 & 0 & 0 & 0 & 0 & 1 & 1 & 0 \\ 
	0 & 0 & 0 & 0 & 0 & 0 & 0 & 1 & 0 & 0 & 0 & 0 & 0 & 0 & 0 & 0 & 0 & 0 & 0 & 1 & 1 \\ 
	0 & 0 & 0 & 0 & 0 & 0 & 0 & 0 & 1 & 0 & 0 & 0 & 0 & 0 & 0 & 0 & 0 & 0 & 1 & 1 & 1
	\\
	\hline
	0 & 0 & 0 & 0 & 0 & 0 & 0 & 0 & 0 & 1 & 0 & 0 & 0 & 0 & 0 & 0 & 0 & 0 & 0 & 1 & 1 \\ 
	0 & 0 & 0 & 0 & 0 & 0 & 0 & 0 & 0 & 0 & 1 & 0 & 0 & 0 & 0 & 0 & 0 & 0 & 1 & 1 & 1 \\ 
	0 & 0 & 0 & 0 & 0 & 0 & 0 & 0 & 0 & 0 & 0 & 1 & 0 & 0 & 0 & 0 & 0 & 0 & 1 & 0 & 1
	\\
	\hline
	0 & 0 & 0 & 0 & 0 & 0 & 0 & 0 & 0 & 0 & 0 & 0 & 1 & 0 & 0 & 0 & 0 & 0 & 1 & 1 & 1 \\ 
	0 & 0 & 0 & 0 & 0 & 0 & 0 & 0 & 0 & 0 & 0 & 0 & 0 & 1 & 0 & 0 & 0 & 0 & 1 & 0 & 1 \\ 
	0 & 0 & 0 & 0 & 0 & 0 & 0 & 0 & 0 & 0 & 0 & 0 & 0 & 0 & 1 & 0 & 0 & 0 & 1 & 0 & 0
	
	\\
	\hline
	0 & 0 & 0 & 0 & 0 & 0 & 0 & 0 & 0 & 0 & 0 & 0 & 0 & 0 & 0 & 1 & 0 & 0 & 1 & 0 & 1 \\ 
	0 & 0 & 0 & 0 & 0 & 0 & 0 & 0 & 0 & 0 & 0 & 0 & 0 & 0 & 0 & 0 & 1 & 0 & 1 & 0 & 0 \\ 
	0 & 0 & 0 & 0 & 0 & 0 & 0 & 0 & 0 & 0 & 0 & 0 & 0 & 0 & 0 & 0 & 0 & 1 & 0 & 1 & 0
	\end{array} \right) 
	$$

	with
	$\mathbf{M}_{0}=\left( \begin{array}{ccc}
	0 & 0 & 0 \\ 
	0 & 0 & 0 \\ 
	0 & 0 & 0
	\end{array} \right), \ \ \ 
	\mathbf{M}_{1}=\left( \begin{array}{ccc}
	1 & 0 & 0 \\ 
	0 & 1 & 0 \\ 
	0 & 0 & 1
	\end{array} \right)\ \ \  and \ \ \ \mathbf{M}_{\alpha^i}=\left( \begin{array}{ccc}
	0 & 1 & 0 \\ 
	0 & 0 & 1 \\ 
	1 & 1 & 0
	\end{array} \right)^i
	$  for all $i\in \left\lbrace 1, 2, ..., 6\right\rbrace$ 
\end{example}

The parity check matrix of the binary image $Im_2(\mathcal{C})$ of the code $\mathcal{C}$ is given by:
$$
\mathcal{H}_2=\left( \begin{array}{ccc|ccc|ccc|ccc|ccc|ccc|ccc}
0 & 0 & 1 & 0 & 1 & 0 & 1 & 0 & 1 & 0 & 1 & 1 & 1 & 1 & 1 & 1 & 1 & 0 & 1 & 0 & 0 \\ 
1 & 0 & 1 & 0 & 1 & 1 & 1 & 1 & 1 & 1 & 1 & 0 & 1 & 0 & 0 & 0 & 0 & 1 & 0 & 1 & 0 \\ 
0 & 1 & 0 & 1 & 0 & 1 & 0 & 1 & 1 & 1 & 1 & 1 & 1 & 1 & 0 & 1 & 0 & 0 & 0 & 0 & 1
\end{array} \right) 
$$ 

Let $u=(2,3,3,2,2,3,3)\in \left\lbrace 1, 2, 3 \right\rbrace ^7$ be a tuple then 
$I_u=\left\lbrace 2, 6, 9, 11, 14, 18, 21\right\rbrace $  Now we compute $S_u\left( \mathcal{H}_2 \right) $ 
corresponding to the generator matrix of $Im_2(\mathcal{C})$:
$$
S_u\left( \mathcal{H}_2\right)=\left( \begin{array}{ccccccc}
0 & 0 & 1 & 1 & 1 & 0 & 0 \\ 
0 & 1 & 1 & 1 & 0 & 1 & 0 \\ 
1 & 1 & 1 & 1 & 1 & 0 & 1
\end{array} \right) 
$$

The generator matrix $\mathcal{G}_{\mathcal{S}_u}$ of the subfield subcode $\mathcal{S}_u(\mathcal{C})$ of 
the code $\mathcal{C}$ over $\mathbb{F}_2$ is given by
$$
\mathcal{G}_{\mathcal{S}_u}=\left( \begin{array}{ccccccc}
1 & 0 & 0 & 0 & 0 & 0 & 1 \\ 
0 & 1 & 0 & 0 & 0 & 1 & 1 \\ 
0 & 0 & 1 & 0 & 1 & 1 & 0 \\ 
0 & 0 & 0 & 1 & 1 & 1 & 0
\end{array} \right) 
$$

Then $\mathcal{S}_u(\mathcal{C})$ is a $[7, 4, 2]$ binary linear code.

When $u=(1, 3, 1,2,3, 1,3)$
we have $I_u=\left\lbrace 1, 6, 7, 11, 15, 16, 21 \right\rbrace$ and the subfield subcode j
$\mathcal{S}_u(\mathcal{C})$ over $\mathbb{F}_2$ of the code $\mathcal{C}$ is an $[7, 4, 3]$ binary linear
code of generator matrix
$$
\mathcal{G}_{\mathcal{S}_u}=\left( \begin{array}{ccccccc}
1 & 0 & 0 & 1 & 0 & 1 & 0 \\ 
0 & 1 & 0 & 1 & 0 & 1 & 1 \\ 
0 & 0 & 1 & 1 & 0 & 0 & 1 \\ 
0 & 0 & 0 & 0 & 1 & 1 & 1
\end{array} \right) 
$$ 
Algorithm \ref{algo:1} give a simple method to construct a generator matrix $G$ of $S_u(\Cc)$ from a 
generator matrix $\Gc$ of $\Cc$.

\begin{algorithm}  
\caption{Generator matrix of $G$ of  $S_u(\Cc)$\label{algo:1}}	
    \begin{itemize}
      \item  Construct a generator matrix of $Im_q(\Cc)$ using the method described in Section 
      \ref{section:qary image}.
   
      \item  Compute a generator matrix of the dual of this image.
   
      \item  Delete the columns indexed by $J_u$ and  perform a Gaussian elimination on this matrix.
      
      \item Compute a generator matrix of the dual of this punctured code.
   \end{itemize}  
\end{algorithm}

\subsection{Subfield Subcode as shortened $q$-ary image of a code }

If we choose a basis $\mathcal{B}=\left\lbrace b_1, b_2, ..., b_m \right\rbrace$ of $\F_{q^m}$  such that 
$b_1=1$, then we have 
$\F_q=\phi^{-1}\left( \left\lbrace (a_1, 0, 0, ..., 0) \: |\: a_1\in\F_q\right\rbrace\right)$.

\begin{proposition} Let $u=(1,1,...,1)$ be a $n$-tuples of positions.  If $\Bc$ is a basis such that $b_1=1$,
then the code $S_u(\Cc)$ 
   is the subfield subcode of $\Cc$ over $\F_q$.
\end{proposition}

\begin{IEEEproof}
   This is a direct consequence of Remark \ref{rem:Racc}: a codeword $c=(c_1,...,c_n)\in \F_{q}^n$ is in  
   $S_u(\Cc)$ if and only if $c'=(c_1,0,...,0,c_2,0,...,0,....,c_n,0,...,0) \in \F_{q}^{nm}$ is in 
   $Im_q(\Cc)$, which is equivalent to the fact that $\Phi^{-1}_{\Bc}(c') \in \Cc \cap \F_q^n$.
\end{IEEEproof}

A natural question is: Does this construction allow to construct new codes? The answer will be given in the
next section.
  
\begin{proposition} Let $u=(i,i,...,i)$ be a $n$-tuples of positions where $1\leq i \leq m$.  If $\Bc$ is a
multiplicative 
   basis of the form $\Bc=(1,\alpha,\alpha^2,...,\alpha^{m-1})$, then the code $S_u(\Cc)$ 
   is the subfield subcode over $\F_q$ of $\Cc$.
\end{proposition}

We do not the give the proof of this result, but it will be easily derived from the discussion of the next 
section.
 
\section{A first attempt at generalization}\label{sec: 3}

In this section we will show that the codes introduced in Section \ref{sec:first-construction} can also be
constructed using the classical
method used to construct alternant codes. However, our approach leads to a second generalization presented in
Section
\ref{sec:subspacesubcode}.

For an arbitrary finite field $\F_q$ we denote by $\E=\F_q^m$ and by  $GL_q(m)$ the linear group of
isomorphisms acting on  $\E$.

\subsection{$q$-ary block codes of length $r$ over $\E$}

For more details on block codes, the reader can refer to \cite{TB}.

\begin{definition}
Let $(A, +)$ be an additive group. An additive code of length $r$ over $A$ is an additive subgroup of
$(A^n,+)$.
\end{definition}

\begin{definition}
A block code of length $n$ over $\E=\F^m_q$ is an additive code over the additive group $(\E^n, +)$ which is 
stable by scalar multiplication of any element $\lambda$ of $\F_q$. The
integer $m$ is the size of the blocks. 
\end{definition}

Note that the condition on the scalar multiplication is not necessary if $q=p$ is a prime number.
Since $\E^n$ is a $\F_q$-linear vector space of dimension $nm$ isomorphic to $\F_q^{nm}$, a block code is
also 
a $\F_q$-linear code of length $nm$. However, in this paper we are not interested in its properties as code
of length $nm$, but in its block properties.

In particular, we look at its block-weight $w_{m}$, which denotes the number of non-zero blocks.
For instance, the $q$-ary image $Im_q(\Cc)$ introduced previously is nothing else than a block code of size of
blocks $m$ and minimum block-distance equal to the minimum distance of $\Cc$.

Since a block code $C$ is a $\F_q$-linear code, it is possible to define the notion of generator matrix, 
which is nothing else than the generator matrix of the corresponding linear code of length $mn$ over $\F_q$. 
If its dimension is $k$, in order to compare a block code of block-size $m$ with a $\F_{q^m}$-linear code, we 
introduce the notion of pseudo-dimension, which is $k/m$. Even if it is possible to construct the $\F_q$-dual
of the 
linear code of length $nm$, the notion of duality for block code is not completely obvious, for 
example the $q$-ary image of the dual of a code $\Cc$ over  $\F_{q^m}$ is not the dual of its   $q$-ary
image. 
More details on additive block codes, some generalizations of generator matrices and a notion of 
block-duality can be found in \cite{TB}.

\subsection{Linear isometries of block codes}\label{sec:isometries}

Generalizing the results of Section \ref{automorphism}, it is possible to define the isometry group of 
$\E^n$. We denote by $w_m(x)$ the block weight of a word $x \in E^n$.

Let $GL_q(m)$ be the group of $\F_q$-linear automorphism. Then  $GL_q(m)$ is isomorphic to the group of
non-singular square 
matrices of length $m\times m$ over $\F_q$. 

For all $f=(f_1, f_2,...,f_n)\in GL_q(m)^n$ and $x=(x_1|x_2|...|x_n)\in \E^n$ where $x_i\in \E$, 
we define the action of $GL_q(m)^n$ on $E^n$ as follows:
$f (x)=(f_1(x_1)|f_2(x_2)|..|f_n(x_n))$.
This is equivalent to multiplying $x$ by the block-diagonal matrix  of size $nm$ whose blocks are matrices of
$f_i$
 
\begin{theorem}\label{isometry}
  The $\F_q$-isometries of $\E^n$ (\ie linear isomorphisms preserving the Hamming block-weight) form a group
generated by block
permutations and $GL_q(m)^n$.
\end{theorem}

\begin{IEEEproof}
Let $Mon_n(GLq(m))$ be the group generated by the block permutations where each block is of length $m$ and
the block diagonal 
matrices of length $mn$ which each matrix of length $m$ on the diagonal is a non-singular matrix. It is clear
that block permutations 
and the block diagonal matrices preserve the Hamming block weight of the 
element of $\E^n$.

 Reciprocally, let $g$ be an isometry of $\E^n$.  We look at the images of elements of $\E^n$ of block weight 
 $1$ by $g$.
 For $1\leq i\leq n$, let $V_i$ be the subspace of $E^n$ of elements with all bock component 
equal to 0 except the $i$-th: if $x \in V_i$, then $x=(0,...,0,x_i,0,...,0)$, $x_i \in E$. Pick an element $x
\in V_i$. 
Since $g$ is a block isometry, $y=g(x)\in V_j$ for some $j$, $1\leq j \leq n$. Suppose that there exists
another element $x' \in  V_i$ such that $g(x') \in V_{j'}$, with $j\neq j'$. Clearly $w_m(x+x')=1$ and
$w_m(g(x+x'))=2$. This implies that $g(V_i)=V_j$. So, $g$ acts as a permutation on the set of $V_i$, which
define the block-permutation part of our isometry. Applying the inverse of this permutation to $g$, we can
now suppose that, for all $i$, $g(V_i)=V_i$. If $g_i$ denotes the restriction of $g$ to $V_i$, $g_i$ must be
$\F_q$-linear, moreover, since $g_i$ preserves the block weight, $Ker(g_i)=\{0\}$, so $g_i \in GL_q(m)$,
which completes the proof of this theorem.

\end{IEEEproof}
 
The ``monomial group'' $Mon_n(GLq(m))$ introduced in the proof of Theorem \ref{isometry}  consists of the
$n\times n$ matrices having one and only one nonzero elements on each row and each column, moreover this
non-zero element must be invertible and then is an element of $GL_q(m)^n$. So, this theorem is a
generalization 
of Section \ref{automorphism}.

\begin{definition}
   Let $C$ and $C'$ be two block codes of length $n$ over $E$. The codes $C$ and $C'$ are equivalent if there 
   exists an element $f\in Mon_n(GLq(m))$ such that $C'=f(C)$.
\end{definition}

Clearly, if $C'=f(C)$, then the minimum block-distances of $C$ and $C'$ are equal.  
Moreover, if there exists a block-distance decoding algorithm for $C$, it can be used to decode $C'$.

There is no natural notion of duality for the block structure of a $F_q$-linear code over $E^n$. 
However, we  can look at the dual of a block code $C$ considered as a code of length $nm$ over $\F_q$.

If $f_i \in GL_q(m)$ is a linear isomorphism, we denote by $f_i^T$ its adjoint isomorphism. 
From a matrix point of view, this means that $M_{f_i^T}=M_{f_i}^T$.

\begin{proposition}\label{prop:equiv-dual}
   Let $C$ be an additive code of length $n$ over $E$, $f=(f_1,...,f_n)\in GL_q(m)^n$ be a diagonal isometry
(without  permutation) and $C'=f(C)$. Let $f^*=((f_1^{-1})^T,...,(f_n^{-1})^T) \in GL_q(m)^n$. Then the
relation 
   between the dual of $C$ and the dual of $C'$ is $C'^{\bot}=f^*(C'^{\bot})$.  
\end{proposition}

\begin{IEEEproof}
   If $x=(x_1,...,x_n) \in \E^n$ and $y=(y_1,...,y_n) \in \E^n$, then we have 
   $<x,y>=\sum_{i=1}^n<x_i,y_i>=\sum_{i=1}^n x_i y_i^T$.
   
   Applying this property to $f(x)$ and $f^*(y)$, we obtain 
   \\ $<f(x),f^*(y)>= \sum_{i=1}^{n}x_iM_{f_i} (y_i 
   (M_{f_i}^{-1})^T)^T=\sum_{i=1}^{n}x_i M_{f_i}M_{f_i}^{-1}y_i^T=\sum_{i=1}^{n}x_iy_i^T=<x,y>$.
   
   Consequently, we have $<x,y>=0$ if and only if $<f(x),f^*(y)>=0$, which completes the proof.
\end{IEEEproof}

In addition, it is easy to verify that the dual of a permuted block code is the permuted block code of its 
dual.

\subsection{Generalized Subfield Subcodes} \label{sec gss}

Combining the results of Section \ref{sec:first-construction} and Section \ref{sec:isometries}, we are able 
to define the notion of generalized subfield subcode of a linear code $\Cc$ over $\F_{q^m}$.

As before, $\Bc$ denotes a fixed basis of $\F_{q^m}$ over $\F_q$, $u$ denotes a set of indexes used to 
construct a shortened code, $f=(f_1,...,f_n) \in GL_q(m)^n$ an $n$-tuple of linear isomorphisms, $\pi$ a 
permutation of the $n$ blocks and $mon=\pi \circ f \in Mon_n(GLq(m))$ the corresponding isometry.

\begin{definition}
Let $\Cc$ be a $\F_{q^m}$-linear code of length $n$. The Generalized Subfield Subcode with relative to $\Bc$, 
$u$ and $mon$ is the $\F_q$-linear code  $GSS(\Cc)=S_{u}(mon(Im_q(\Cc)))$.
\end{definition}

We can immediately make some remarks.

\begin{remark} \mbox{ }
   \begin{itemize}
      \item  If $\Cc$ is an $[n,k,d]_{q^m}$-linear code, then $GSS(\Cc)$ is an 
      $[n, k' \geq n-m(n-k), d' \geq d]_q$-linear code. Moreover, if $\Cc$ has a decoding algorithm up to $t$
errors, 
      this algorithm can be applied to  $GSS(\Cc)$.

      \item  In order to construct all the Generalized Subfield Subcodes of a code $\Cc$, it is not necessary
to 
      change the basis $\Bc$, indeed, a  change of basis can be made by applying the corresponding matrix to 
      the coordinates of $f$.
   
      \item  In order to construct all the Generalized Subfield Subcodes of a code $\Cc$, it is sufficient to 
      use the projections indexed by $\bar{1}=(1,1,...,1)$. Indeed, other values for the coordinates$u_i$ 
      correspond to permutations on each $m$-blocks, which is always a linear mapping in $GL_q(m)$ and can be 
      composed with the $f_i$'s.

   \end{itemize}
\end{remark}

Following Algorithm  \ref{algo:1}, Algorithm  \ref{algo:2} allows to construct a generator matrix of a GSS code. The
basis $\Bc$ is fixed and $u=(1,...,1)$. 

\begin{algorithm} 
 \caption{Generator matrix of a GSS code \label{algo:2}}	

  {\em\textbf{Input:}} A generator matrix $\Gc$ of $\Cc$ and $mon\in Mon_n(GLq(m))$

   {\em\textbf{Output:}} A generator matrix $G$ of $GSS(\Cc)$, relative to $mon$.
   
    \begin{enumerate}
      \item  Construct a generator matrix $M$ of $Im_q(\Cc)$
      
      \item Compute $M'=M Diag_{f}$ where $Diag_{f}$  is the $nm \times nm$ block diagonal matrix, with each 
      block corresponding to the $f_i$'s.
   
      \item  Compute a parity check matrix $H'$ of $M'$.
   
      \item  Delete the columns of the matrix $H'$ except the first ones of each block. This leads to a 
      parity check matrix $H$ of $GSS(\Cc)$. 
      
      \item Perform a Gaussian elimination on $H$.
     
      \item Permute $H$ according to $\pi$.
        
      \item Compute a parity check matrix $G$ of $H$
   \end{enumerate}  

   {\em\textbf{Return:}} $G$

\end{algorithm}

A first remark is the fact that the permutation $\pi$ can be applied at any time from step 3 in the 
algorithm. However, it is simplest to perform the permutation at the end, since we no longer have to apply 
this permutation on blocks, but only on vectors of length $n$.

In addition, it is possible to use Proposition $\ref{prop:equiv-dual}$ in the algorithm by inverting the 
order of Step 2 and Step 3 and replacing $f$ by $f^*$. This give the following variant for steps 2) to 4):

\begin{description}
   \item[2)]  Compute a parity check matrix $H$ of $M$.

   \item[3)]  Compute $H'=H Diag_{f^*}$.

   \item[4)]  Delete the columns of the matrix $H'$ except the first ones of each block. This leads to a 
      parity check matrix $H$ of $GSS(\Cc)$. 
\end{description}

Let $p_1$: $E \mapsto \F_q$ be the projection of an element on to its first component, the operations 2) and
3) 
of this variant can be combined into a single map $p_1(f^*)=(p_1\circ (f_1^{-1})^T,...,p_1\circ (f_n^{-1})^T) 
\in (E^*)^n$ where $E^*$ is the dual vector space of $E$, \ie $E^*=\Lc(E,\F_q)$.

Remember that $E^*$ is isomorphic to $E$ as follows: for $y \in E$, we denote by $\phi_y \in \E^*$ the map 
defined by $\phi_y(x)=<x,y>=x y^T$. So, instead of choosing an element $mon=\pi \circ  f$ as input of 
Algorithm \ref{algo:2}, we can choose a permutation $\pi$ and an $n$-tuple $y=(y_1,...,y_n) \in (E\setminus 
\{0\})^n$.   We denote by $Diag_y$ the $nm \times n$ block diagonal matrix with diagonal blocks $y_i^T$. Note 
that the diagonal blocks are not square matrices, but column vectors. So, the mapping $p_1(f^*)=y$ is 
computed using $Diag_y$: for $x=(x_1,...,x_n) \in E^n$, 
$y(x)=(x_1 y_1^T,...,x_ny_n^T)= x Diag_y$. This leads to Algorithm   \ref{algo:3}:

\begin{algorithm} 
  \caption{Generator matrix $G$ of $GSS(\Cc)$ relative to $\pi$ and $y$ \label{algo:3}}
   {\em\textbf{Input:}} A generator matrix $\Gc$ of $\Cc$, a permutation $\pi$, and a matrix $Diag_y$ with 
   $y_i \neq 0$ for $1\leq i\leq n$.
   
   {\em\textbf{Output:}} A generator matrix $G$ of $GSS(\Cc)$, relative to $\pi$ and $y$.
   
    \begin{enumerate}
      \item  Construct a generator matrix $M$ of $Im_q(\Cc)$
      
    \item  Compute a parity check matrix $H$ of $M$.
   
      \item  Compute $H'=M Diag_y$.

      \item Perform a Gaussian elimination on $H'$.
     
      \item Permute $H'$ according to $\pi$.
        
      \item Compute a parity check matrix $G$ of $H'$
   \end{enumerate}  

   {\em\textbf{Return:}} $G$
\end{algorithm}

\subsection{Link between generalized subfield subcodes and subfield subcodes of equivalent codes}
\label{sec:V.D}

In this section, we show that the generalized subfield subcodes of a given code are nothing else than
subfield 
subcodes of equivalent codes. However, the approach presented in Section \ref{sec 2} gives a new 
point of view on this topic and will naturally be extended in the next section.

We need to have a more algebraic approach of the construction of generalized subfield subcodes. Suppose 
first without loss of generality, that the block permutation $\pi$ is the identity. Indeed, this permutation
can 
always be considered as having already been applied beforehand to the code $\Cc$.

We will look at a fixed coordinate of a word $c=(c_1,...,c_n)\in \Cc \subset \F_{q^m}^n$. 
We choose a coordinate $u=c_i \in \F_{q^m}$. Suppose that $u=\sum_{i=1}^{m}u_i b_i$, $u_i \in \F_q$ is the
decomposition 
of $u$ on the basis $\Bc$. Let $M=M_{f_i}$ be a $m \times m$ matrix corresponding to $f_i \in GL_q(m)$. 
This matrix $M$ can be interpreted as a change of basis $\Bc$ to $\Bc'=(b'_1,...,b'_m)$: $(u_1,...,u_m)M$ is 
nothing else than the coordinates of $u$ on this new basis $\Bc'$. Let $V_i=V$ be the $\F_q$-subspace of 
$\F_{q^m}$ generated by $b'_1$. The shortening operation in the $i$-th $m$-tuple in the construction of a 
generalized subfield subcode  consists of keeping only the code words having their $i$-th coordinate in 
$V_i$, and then identify $V_i$ to $\F_q$by means of its generator $b'_1$.

We have shown the following proposition:

\begin{proposition}\label{prop:subspace}
   The generalized subfield subcodes of a $\F_{q^m}$-linear code $\Cc$ can be constructed as follows:
   \begin{enumerate}
      \item  Choose a set of $n$ $\F_q$-subspaces $V_i$ of rank 1 of  $\F_{q^m}$.
   
      \item  Set $\Cc'=\Cc \cap \prod_{i=1}^n V_i$.
   
      \item  By means of a generator $a_i$ of each $V_i$, identifies $V_i$ to $\F_q$. This leads to a $q$-ary 
      image $C=Im_q(\Cc')$.
      
      \item Choose a permutation $\pi$ over $\F_q^n$ and return $\pi(C)$.
   \end{enumerate}
\end{proposition}

One can remark that, since $C$ is a $\F_q$-linear code, the construction does not depend on the choice of 
representatives $a_i$.

As a consequence of Proposition \ref{prop:subspace}, we obtain the following theorem:

\begin{theorem}\label{th:equiv}
   Let $\Cc$  be an $[n,k]_{q^m}$-linear code. The generalized subfield subcodes of $\Cc$ are exactly the 
   codes obtained by taking the subfield subcodes of the $\F_{q^m}$-linear codes equivalent to $\Cc$ under 
   $\F_{q^m}$-linear isometries (as described in Section \ref{automorphism}).
\end{theorem}

\begin{IEEEproof}

Without lost of generality, we can suppose that, for both the generalized subfield subcode construction and 
the subfield subcode of equivalent codes construction, the permutation $\pi$ is the identity. 

Note that the subfield subcode of $\Cc$ over $\F_q$ corresponds to the choice $V_1=...=V_n=\Lc(1)=\F_q 
\subset \F_{q^m}$.

Consider now any choice of subspaces $V_i=\Lc(a_i)$. Set $D=Diag(a_1,...,a_n)$ be the $n \times n$ diagonal 
matrix which corresponds to the multiplication of each component by the $a_i$'s. The subfield subcode of the 
image of $\Cc$ by $D$ is clearly the generalized subfield subcode of $\Cc$ corresponding to $\prod_{i=1}^n 
V_i$.   
\end{IEEEproof}

The following corollary is a direct consequence of Theorem \ref{th:equiv}.

\begin{corollary}
    The Generalized Subfield Subcodes of Reed-Solomon codes are exactly  Alternant codes.
\end{corollary}

In addition, we will make explicit the link between the subspaces $V_i$ of Proposition \ref{prop:subspace}
and the $y_i$'s of 
Algorithm \ref{algo:3}.

\begin{proposition}
   The vector spaces of  Proposition \ref{prop:subspace} are generated by the elements $y_i$ of Algorithm
\ref{algo:3}. 
\end{proposition}

\begin{IEEEproof}
    As previously, we denote by $M=M_{f_i}$ the $m \times m$ matrix corresponding to $f_i \in GL_q(m)$. This 
matrix $M$ is interpreted as a change of basis $\Bc$ to $\Bc'=(b'_1,...,b'_m)$. The matrix $M^{-1}$ 
corresponds to the change of basis from $\Bc'$ to $\Bc$. Its first row is given by the coordinates of $b'_1$
in the 
basis $\Bc$. In the construction of Algorithm  \ref{algo:3}, the coordinates of $y_i$ are given by the first
column of
$(M^{-1})^T$. 
Consequently, we have $b'_1=a_i=y_i$, which completes our proof.
\end{IEEEproof}

\section{Subspace subcodes}\label{sec:subspacesubcode}

Codes defined over a finite field of great size are used to 
correct burst errors or for concatenation of codes. The most famous example is that of Reed-Solomon codes 
over $\F_{2^m}$, with typical values $4 \leq m \leq 8$. However Reed-Solomon codes are MDS codes, which 
implies in particular that their length $n$ is limited to $2^m$.

In practice, for transmission applications, a code over $\F_{2^m}$ is generally implemented in binary, \ie 
using its binary image $Im_2(\Cc)$. The notion of additive block codes over $E=\F_2^m$  is an interesting and 
efficient alternative for applications. In this section, we will present a generic construction of additive 
block codes of length greater than $2^m$. If the starting code is a Reed-Solomon code, these new codes posses 
a decoding algorithm and have a constructed minimal distance, which remains very competitive even if these 
codes cannot be MDS.

In this section, we introduce a new class of additive block codes with interesting parameters for both 
transmission and cryptology applications.

In order to facilitate a comparison between the parameters of  linear codes over $\F_{q^m}$ and block 
codes over $\F_{q}^m$, we use the notation $[n;k;d]_{q^m}$ for their parameters, with $n$ is the $m$-block
length 
of the code, $k =Log_{q^m}({}\sharp C)$ is its pseudo-dimension and $d$ is its $m$-block minimum distance. 
Note that, for an additive block code, $k$ is not necessarily an integer.

In order to simplify the presentation of this section, we do not discuss the presence of a possible 
permutation $\pi$ which is implicitly fixed to be the identity.

\subsection{Subspace subcodes}

A natural, simple and efficient way to generalize the approach introduced in Section \ref{sec:V.D} is to 
increase the size of the subspaces $V_i$.

\begin{definition}
   Let $\Cc$ be an $\F_{q^m}$-linear code of length $n$ and $V$  be a $\F_q$-subspace of $\F_{q^m}$ of 
   dimension $r\leq m$. The subspace subcode over 
   $V$ of $\Cc$ is the $\F_q$-linear code $SS_{V}(\Cc)=\Cc \cap V^n$.
\end{definition}

Most of the previous results can be generalized directly. Fixing a basis 
$\Bs=(\beta_1,...,\beta_r)$ of $V$, the code $SS_{V}(\Cc)$ can be identified by an additive block code over
$\Ec=\F_q^r$. If we complete the basis $\Bs$ into a basis $\Bc$ of $E$, this block code is obtained by 
shortening the $q$-ary image of $\Cc$ over the $m-r$ last components of each block.

We deduce directly the following proposition:

\begin{proposition}\label{prop:parameters}
    If the parameters of $\Cc$ are $[n,k,d]_{q^m}$, then those of $SS_{V}(\Cc)$ are 
   $[n,k'\geq (km-n(m-r),d'\geq d]_{q^r}$. 
\end{proposition}

Note that, if we choose another basis $\Bs$ of $V$, it leads to an equivalent (in the meaning of Section 
\ref{sec:isometries}) block code.

In addition, if there is a decoding algorithm for $\Cc$, this algorithm can be applied to decode
$SS_{V}(\Cc)$.

\subsection{Generalized subspace subcode}

\begin{definition}
   Let $\Cc$ be a $\F_{q^m}$-linear code of parameters $[n;k;d]_{q^m}$. Let $r$ be an integer less than $m$. 
   Let $V_1$, ..., $V_n$ be a set of $n $ $\F_q$-subspaces of $\F_{q^m}$ of dimension $r$. Set 
   $W=\prod_{i=1}^{n}V_i$, constituted of $n$-tuples with the $i$-th coordinate in $V_i$. 
   The generalized subspace subcode of $\Cc$ relative to $W$ is the $\F_q$-vector 
   space $GSS_{W}(\Cc)=\Cc \cap W$.
\end{definition}

\textit{Proposition} \ref{prop:parameters} can be also applied to generalized subspace subcodes.

\begin{example}
   Following Example \ref{example:1}, we start from the Reed Solomon code $RS_{3}$ with parameters $[7;5;3]_8$.
   A generator matrix of the dual of its binary image is
   
   $$\Hc=\left(
   \begin{array}
   {ccc|ccc|ccc|ccc|ccc|ccc|ccc}
1 & 0 & 0 & 0 & 0 & 0 & 1 & 0 & 1 & 1 & 0 & 0 & 1 & 0 & 1 & 0 & 0 & 1 & 0 & 0 & 1 \\
0 & 1 & 0 & 0 & 0 & 0 & 1 & 1 & 1 & 0 & 1 & 0 & 1 & 1 & 1 & 1 & 0 & 1 & 1 & 0 & 1 \\
0 & 0 & 1 & 0 & 0 & 0 & 0 & 1 & 1 & 0 & 0 & 1 & 0 & 1 & 1 & 0 & 1 & 0 & 0 & 1 & 0 \\
0 & 0 & 0 & 1 & 0 & 0 & 0 & 1 & 1 & 1 & 0 & 0 & 1 & 1 & 1 & 1 & 1 & 1 & 0 & 1 & 1 \\
0 & 0 & 0 & 0 & 1 & 0 & 1 & 1 & 0 & 0 & 1 & 0 & 1 & 0 & 0 & 1 & 0 & 0 & 1 & 1 & 0 \\
0 & 0 & 0 & 0 & 0 & 1 & 1 & 1 & 1 & 0 & 0 & 1 & 1 & 1 & 0 & 1 & 1 & 0 & 1 & 1 & 1 \\
   \end{array}
   \right)$$
   
   We choose $W= V_1V_2V_1V_3V_1V_2V_1$, where $V_1$ is generated by $1$ and $\alpha$, $V_2$ by $1$ and 
   $\alpha^2$ and $V_3$ by $\alpha$ and $\alpha^2$. So, in order to obtain a parity check matrix of  
   $\Cc=GSS_W(RS_3)$, we delete the columns indexed by 3, 5, 9, 10, 15 17 
   and 21 of $\Hc$.  A binary generator 
   matrix of $\Cc$ is then 

   $$\Gc=
\left(
 \begin{array}
   {cc|cc|cc|cc|cc|cc|cc|cc|c|cc|cc}
1 & 0 & 0 & 0 & 0 & 0 & 0 & 0 & 1 & 0 & 1 & 0 & 0 & 0 \\
0 & 1 & 0 & 0 & 0 & 0 & 0 & 0 & 0 & 1 & 0 & 1 & 0 & 0 \\
0 & 0 & 1 & 0 & 0 & 0 & 0 & 0 & 0 & 0 & 1 & 0 & 1 & 0 \\
0 & 0 & 0 & 1 & 0 & 0 & 0 & 0 & 0 & 0 & 0 & 1 & 0 & 1 \\
0 & 0 & 0 & 0 & 1 & 0 & 0 & 0 & 1 & 0 & 1 & 0 & 1 & 0 \\
0 & 0 & 0 & 0 & 0 & 1 & 0 & 0 & 0 & 1 & 0 & 1 & 0 & 1 \\
0 & 0 & 0 & 0 & 0 & 0 & 1 & 0 & 1 & 0 & 0 & 0 & 1 & 0 \\
0 & 0 & 0 & 0 & 0 & 0 & 0 & 1 & 0 & 1 & 0 & 0 & 0 & 1 \\

\end{array}
\right)$$
   
As a binary code, its parameters are $[14;8;3]_2$. If we look at this code as a block-code of block size 
equals to 2, it is a $[7;4;3]_4$ code, which is optimal compared to a code over $\F_4$ of length 7 and 
dimension 4.

\end{example}

In practice, if we want to construct such codes, it is easy to extend the previous algorithms
to Algorithm \ref{algo:4}.

\begin{algorithm}    \mbox{ } 
 \caption{Generator matrix $G$ $G$ of $GSS_{W}(\Cc)$\label{algo:4}}
 
   {\em\textbf{Input:}} A generator matrix $\Gc$ of $\Cc$, a set $W=\prod_{i=1}^{n}V_i$ of $n$ vector spaces 
   $V_i$ of dimension $r$. Each $V_i$ is defined by a basis $(v_{i,1},...,v_{i,r})$ of $V_i$.
   
   {\em\textbf{Output:}} A generator matrix $G$ of $GSS_{W}(\Cc)$.
   
    \begin{enumerate}
      \item Construct a generator matrix $M$ of $Im_q(\Cc)$
      
    \item  Construct $n$ matrices $M_i$ of size $r \times m$. The rows of $M_i$ are the coordinates of 
    the basis $(v_{i,1},...,v_{i,r})$ of $V_i$ relative to $\Bc$.
   
      \item  Compute $D=Diag(M_1^T,...,M_n^T)$, the block-diagonal matrix of size $mn \times rn$ having the 
      matrices $M_i^T$ as diagonal blocks.
      
      \item Compute $H'=HD$.
      
      \item Perform a Gaussian elimination on $H'$.
         
      \item Compute a parity check matrix $G$ of $H'$
   \end{enumerate}  

   {\em\textbf{Return:}} $G$
\end{algorithm}

\subsection{Examples}\label{sec:example}

In this construction, the minimum distance of the original code is preserved, while the dimension of the code
decreases 
with the number of punctured positions. So, the value $r=m-1$ seems interesting to provide codes 
with nice parameters.

We give some examples.

\begin{itemize}
   \item  $q=2$, $m=4$, $r=3$. We start from the extended Reed-Solomon code $\Cc$ over $\F_{16}$ with
parameters 
   $[16;13;4]_{16}$. For $r=3$, the parameters of  any generalized subspace subcode of $\Cc$ are $[16;k'\geq 
   12;d'\geq 4]_{8}$. Note that the parameters $[16;12;4]_{8}$ are optimal for $\F_8$-linear code.
   
In practice, all the codes we obtained had parameters exactly $[16;12;4]_{8}$.

   \item  $q=2$, $m=5$, $r=3$. We start from the extended Reed-Solomon code $\Cc$ over $\F_{32}$ with
parameters 
   $[32,26,7]_{16}$. For $r=3$, the parameters of  any generalized subspace subcode of $\Cc$ are $[32;k'\geq 
   22;d'\geq 7]_{8}$. The parameters $[32;22;7]_{8}$ are optimal for $\F_8$-linear code.
   
   \item  $q=2$, $m=9$,. We start from the extended Reed-Solomon code $\Cc$ over $\F_{2^9}$ with parameters 
   $[512,350,163]_{512}$. For $r=83$, the parameters of  any generalized subspace subcode of $\Cc$ are
$[512;k'\geq 
   329.75;d'\geq 163]_{256}$. 
   
\end{itemize}

\subsection{Cryptographic applications}

The purpose of this section is not to present the complete design of a public key cryptosystem, but to show
that the generalized subspace subcode construction is a promising research direction to hide the structure of
a code.

The general principle of such a cryptosystem is as follows: the starting point is a class of codes for which 
there exists an efficient decoding algorithm up to a fixed number $t$ of errors. The structure of such a code
is 
masked by some operations which constitute the secret key. The public key is then a generator matrix of a 
code which looks like a random code $C$. The message to be encrypted is encoded by a generator matrix of $C$
and a 
random error of weight $t$ is added to this message.

Such a cryptosystem is sensitive to two types of attacks:
\begin{itemize}
   \item  Structural attacks that involve retrieving the structure of the masked code.

   \item  Decoding by brute force. This consists of applying generic decoding algorithms to a random code.
This 
   problem is NP-hard, however the parameters of the code must be sufficiently large to resist at this kind 
   of attacks.
\end{itemize}

The evaluation of the brute force decoding attack is not easy, and many papers are devoted to this topic
\cite{BernCo,FinSend,LeeBrick,SendTor}. We chose to use a simple criterion: for a given code of parameters
$[n;k;d]$
with correction capacity $t=\lfloor (d-1)/2\rfloor$, we compute the ratio between the number of information 
sets and the number of information sets without errors. Our measure of the workfactor is then
$wf={\binom{n}{k} }/{\binom{n-t }{k}}$. Our criterion yields a workfactor greater or equal to $2^{128}$.

Most of McEliece-like cryptosystems use subfield subcodes of Generalized Reed-Solomon codes (for example the 
binary Goppa codes in the original McEliece cryptosystem). We propose to use generalized subspace subcodes of 
Reed-Solomon codes (GSS codes of GRS codes for short).

There are some advantages and disadvantages to using generalized subspace subcodes instead of subfield 
subcodes.

\begin{itemize}
   \item  In the case $r=1$ and $q=2$, our generalized subspace subcodes are nothing else than Alternant 
   codes, and it is well-known that Goppa codes have better parameters.

   \item  For $r$ close to $m$ (typically $m-4 \leq r m$), the code parameters become more interesting.
Moreover, if we want to construct subfield subcodes of a code over $\F_{q^m}$, the size of subfield is 
   bounded above by $2^{m'}$, with $m'\leq m/2$. So our parameters are more flexible. Finally, there exist
some attacks against subfield subcodes of GRS codes over large fields \cite{FOPPT,FOPT}. A priori, this 
   type of attack does not apply to generalized subfield subcodes.

   \item  The main reason for this comes from the fact that the GSS 
   codes are no longer defined over a field but over some vector space. In return, the description of these
codes as linear codes over $\F_{2^{m'}}$ cannot be used. We need to give a full $\F_q$ basis of our codes, 
   which increase the size of the secret key.
\end{itemize}

In practice, a binary Goppa code of parameters $[4096;3556;91]$ leads to a resistance against brute force
decoding greater than our criterion $wf\geq 2^{128}$. The corresponding size of the public key is then 938
Ko.

Our third example in Section \ref{sec:example} has for parameters $[512;329;163]_{2^8}$ and leads to a
workfactor greater than $2^{128}$.
In addition, one can notice that we did not take in account the fact that this code is defined over a large
alphabet, which will increase the complexity of this attack.
Unfortunately, the size of the secret key is very large, approximatively 1514 Ko.
This is essentially due to the fact that $q=2$, that implies the code is $\F_2$-linear.

An intermediate solution consists in choosing a relatively large $q$ and a small $m$.
Set $q=2^4=16$, $m=3$ and $r=2$.
We obtain $\F_{q^m}=\F_{2^{12}}$, so it is possible to pick a Reed-Solomon code up to the length
$2^{12}=4096$.
For example, we choose a Reed-Solomon code of parameters $[700;580;121]_{2^{12}}$.
The $V_i$ subspaces of our construction are subspaces over $\F_{16}$ of dimension 2.
We obtain a $\F_{16}$-linear GSS code of parameters $[700;520;121]_{2^8}$, which leads to a workfactor $wf$
greater than $2^{128}$.
The $\F_q$-generator matrix is of size $1400 \times 1040$.
Each entry is over $\F_{2^4}$ and needs 4 bits of memory.
As usual, we use a systematic matrix to describe our code \ie a matrix of size $k \times n-k$.
The size of the public key is then $1040 \times 360 \times 4 = 1497600$ bits, or 183 Ko, which is 
significantly smaller than
a classical Goppa code with the same level of security against brute force decoding.

\section{Conclusion}

The purpose of this paper is not to present a complete study of structural  attacks against subspace
subcodes. 
Here is a list of questions that naturally come to mind and deserve further development.

\begin{enumerate}
	\item  The equivalent of GRS codes in the subspace subcode context correspond to generalized subspace 
	subcodes of Reed-Solomon code with parameter 
	$r=m$. It is well known that, from a generator matrix of a GRS code, it is easy to recover the underlying
	algebraic 
	structure, \ie the support of the corresponding Reed-Solomon code and the values of the scalar 
	multiplications on each component \cite{Sidel'nikov/Shestakov:1992}.
	
	The problem in the GSS context is the following: we fix a Reed-Solomon code $\Cc$ of parameters $[n;k;d]$
	over  $\F_{q^m}$. We choose a basis $\Bc$ and compute $Im_q(\Cc)$. We pick randomly 
	an element $mon$ in $Mon_n(GL_{q}(m))$ and compute a $\F_q$-generator matrix $G$ of $mon(Im_q(\Cc))$. From 
	the matrix $G$, is it possible in reasonable time to recover $RS$ and $mon$ or another equivalent set of 
	parameters $RS'$ and $mon'$?
	
	\item  Given a Reed-Solomon code $\Cc$ and a $\F_q$-generator matrix under systematic form of a 
	generalized subspace subcode $GSS_{W}(\Cc)$, is it possible 
	to recover the secret basis of the subspaces $V_i$ and the permutation $\pi$? This question is connected to
a list of problems in increasing order of difficulty:
	
	\begin{itemize}
		
		\item  Suppose $\pi$ is known (which equivalent to $\pi$ is the identity).
		\begin{itemize}
			\item  $r=m$. This particular case will probably be solved using the conjugacy of matrices.
			
			\item  $1 \leq  r <m$.
		\end{itemize}
		
		\item  $\pi$ is unknown.
		\begin{itemize}
			\item  $r=m$. 
			
			\item  $1 \leq  r <m$.
		\end{itemize}

	\end{itemize}

	\item Given  a $\F_q$-generator matrix under systematic form of a 
	generalized subspace subcode $GSS_{W}(\Cc)$ with $\Cc$ an unknown Reed-Solomon code, is it possible 
	to recover $\Cc$ and the algebraic parameters of $GSS_{W}(\Cc)$?
	
\end{enumerate}

\bibliographystyle{IEEEtran}

\end{document}